# Recent Progress in Chirality Research Using Circularly Polarized Light

**Tsubasa Fukue**

National Astronomical Observatory of Japan,

2-21-1 Osawa, Mitaka, Tokyo 181-8588, Japan

E-mail: tsubasa.fukue@nao.ac.jp

## ABSTRACT

We review recent studies of chirality using circularly polarized light, along with the birth and evolution of life and planetary systems. Terrestrial life consists almost exclusively of one enantiomer, *left-handed* amino acids and *right-handed* sugars. This characteristic feature is called homochirality, whose origin is still unknown. The route to homogeneity of chirality would be connected with the origin and development of life on early Earth along with evolution of the solar system. Detections of enantiomeric excess in several meteorites support the possibility that the seed of life was injected from space, considering the possible destruction and racemization in the perilous environment on early Earth. Circularly polarized light could bring the enantiomeric excess of prebiotic molecules in space. Recent experimental works on photochemistry under ultraviolet circularly polarized light are remarkable. Asymmetric photolysis by circularly polarized light can work for even amino acid leucine in the solid state. Amino acid precursors can be asymmetrically synthesized by circularly polarized light from complex organics. Recent astronomical observations by imaging polarimetry of star-forming regions are now revealing the distribution of circularly polarized light in space. Enantiomeric excess by photochemistry under circularly polarized light would be small. However, several mechanisms for amplification of the excess into almost pure enantiomers have been shown in experiments. When enantiomeric excess of amino acids appears in the prebiotic environment, it might initiate the homochirality of sugars as a catalyst. Astrobiological view on chirality of life would contribute to understanding of the origin and development of life, from the birth and the end of stars and planetary systems in space. Deep insights on terrestrial life, extrasolar life, and the origin of life in the universe, would be brought by consideration of both of the place where life is able to live, *the habitable zone*, and the place where life is able to originate, what we shall call *the originable zone*.

# 1. INTRODUCTION

## 1.1 Astrobiology

There are a lot of astronomical objects that human beings have been observing so far. Findings of new type of celestial bodies have been encouraging us to make a step toward new idea. In the end of $20^{th}$ century, we faced the new historical situation. Mayor and Queloz discovered a planet candidate around a solar-like star, 51 Pegasi, in 1995, for the first time [1]. They observed spectrum of a star, not a planet, and detected the oscillation of the spectrum due to Doppler effect because both of a star and its planet orbit their barycenter. This planet is thought to be one of hot Jupiters. The separation between the host star and the planet is about 1/20 of that between our Sun and Earth, and the mass of the planet is about a half of Jupiter in our solar system. While the terrestrial life seems not to live on this reckless planet, this discovery encouraged us to consider the existence of planets around stars other than our Sun, extrasolar planets (ESPs). Following this success, Charbonneau and co-workers observed a star whose ESP candidate had been indicated by Doppler effect in 2000 [2]. They discovered that the star transiently dims because the planet occults the host star when the planet transits between the host star and us. These detections of both of Doppler effect and transit on a star indicate the existence of ESPs strongly. Direct detection of ESPs has been also tried, suffering from darkness of ESPs very close to the bright host star, uncertainties of determination of mass of detected objects, and determination whether the ESP candidate accompanies the star. In 2008, Kalas and colleagues reported the direct detection of an ESP using Hubble Space Telescope [3]. Remarkably, they compared direct images of an ESP observed in 2004 and 2006, and showed the orbital movement of the ESP. In September 2010, the number of all the ESP candidates is approaching 500. These discoveries in 15 years have impacted us. Now, it would be thought that 8 planets in our solar system and ESPs are not special in space. It is possible that another Earth would exist around another Sun. Astronomers are searching ESPs, and also biomarkers which indicate the existence of extrasolar life [4]. When observational data of a lot of ESPs [5] and their circumstances will be available in future, it would lead to understanding many evolutional stages, from their birth to end, along with life on planets. Astronomical view involves origins of life on ESPs, which might involve the hint of the origin of the life on Earth. Naturally, the new academic field, *Astrobiology*, has been launched in face of the new interdisciplinary stage involving extraterrestrial life and origins of various life including us in space.

**1.2 Habitability and Originability**

The discoveries of ESPs have led to deep consideration of the possible area where life is able to live, *a habitable zone* [6, 7]: a habitable zone in a stellar system [8], a habitable zone in a galaxy [9], a habitable zone in the universe, involving planets [10] and moons [11], and various type of the host star [12, 13]. In view of habitability, the spatial distribution of temperatures, and the spatial distribution of elements for terrestrial planets and terrestrial life, around the host star, along with the evolution of the planetary atmosphere, are sometimes focused on.

However, it is still unknown whether the origin of life is always generated in a habitable zone. It is important to consider not only *a habitable zone*, where life is able to live, but also the place where life is able to originate, what we shall call *the originable zone*. A habitable zone and an originable zone would not be equal. However, they would have partial overlap. When biomarkers of ESPs are directly observed in future, it would be fruitful to consider both of *habitability* and *originability* for life in space, to confirm the biomarker. In particular, there are ESPs around stars whose type is different from our Sun. Generally, we do not know extrasolar life on ESPs, even extrasolar life on earth-type planets around sun-like stars. There might be various type of life, and various route to life in space. In this chapter, originability is focused on, rather than habitability.

**1.3 Enantiomer and Homochirality of the Terrestrial Life**

Life on Earth consists almost exclusively of one enantiomer, *left-handed* amino acids and *right-handed* sugars. Enantiomers are some kind of isomers, involving pair. Atoms of an enantiomer are bonded similarly to that of another enantiomer, and then chemical property of an enantiomer resembles with that of another enantiomer. On the other hand, the spatial configuration of atoms of an enantiomer is set as a mirror image of that of another enantiomer. Enantiomers of amino acids consist of left-handed amino acids and right-handed amino acids. Usual products of amino acids are racemic, which include almost the same quantity of left- and right-handed amino acids. However, the terrestrial life uses almost only left-handed amino acids. This characteristic feature of biomolecular is called *homochirality*, whose origin is still unknown. Some right-handed amino acids work for life, and right-handed amino acids would be involved with aging and disease [14, 15, 16, 17]. Enantiomers are essential for terrestrial life.

**1.4 Early Earth and Early Life**

The route to homogeneity of chirality, which is characteristic of terrestrial life, would be involved with the origin and development of life [18, 19, 20, 21]. The age of our solar system is thought to be about 4.6 billion years, indicated by analysis of meteorites [22]. When we look back at the initiation of life, we would watch primeval Earth and our solar system, where life and its material were formed. Earth is thought to be formed by accumulation of planetesimals and protoplanets [23], as well as these events seem to occur in extrasolar planetary systems [24]. The accumulation is thought to have brought thermal energy from gravitational energy to early Earth, about 4.6 billion years ago. There could be a molten magma ocean on early Earth [25, 26], high temperature, and bombardment from comets and asteroids [19], as well as magma oceans were present on other bodies in the early solar system [27]. In particular, the giant impact [28], the impact between protoplanets accreted from planetesimals [29] which would yield Moon [30, 31], would bring a completely molten magma ocean [32, 33], which would yield the core and mantle of Earth [34, 35]. Such a heat environment would destroy enantiomers, even if homochirality occurred on start of Earth. Therefore, the origin of life with homochirality would be initiated after Earth cooled down, in short time before the first terrestrial life appeared (probably about 3.8 billion years ago) [36, 18, 19]. The mechanism of production of homochirality is very controversial, involving many chemical routes on Earth, and in space [18, 19].

**1.5 The Birth Place of Our Solar System**

In space, low mass stars such as our Sun can be formed in two-type star-forming regions: a massive star-forming region where both high-mass and low-mass stars are formed, or a relatively isolated region where only low-mass stars are formed [37, 38]. The decay products of short-lived radionuclides in meteorites indicate the birth place of our solar system [37]. $^{60}$Fe has a half-life of ~1.49 million years. This half-life is very short, in comparison with the age of our solar system (~4.6 billion years). $^{60}$Fe is a neutron-rich isotope that is formed exclusively in stars, e.g., in core collapse supernova which is the end of a massive star. The presence of $^{60}$Fe in primitive meteorites is confirmed, suggesting that a supernova explosion occurred near our Sun [39, 40, 41, 42]. This indicates that the birth place of our solar system was located in a massive star-forming region. The nearest massive star-forming region is the Orion nebula, and then it is a familiar and important target in astronomy.

## 1.6 Linear and Circular Imaging Polarimetry in Astronomy

Because we can not directly explore astronomical objects far from Earth, it is essential to investigate the property of light from the astronomical objects. The property of light involves wavelength, and polarization [43, 44]. When the light is scattered, the light is often polarized depending on the scattering angle and the optical property of scattering body. In many cases in the universe, the light from stars is scattered at the circumstellar matter and polarized. In particular, during the formation of star/stars and its planetary system, dust grains around the host star/stars evolve and polarize the stellar light. The polarization yields useful information for the circumstellar structures and the property of dust grains.

The polarization status generally consists of linear and circular polarizations, which are obtained with linear and circular polarimetry in astronomy, respectively. In partially polarized light, circular polarization (CP) can not be derived from linear polarization (LP), vice versa. To obtain clear view of polarization status, it is preferable to perform linear and circular polarimetry with the same telescope. The 1.4-m IRSF telescope, which is set up with the SIRIUS camera [45] and its polarimeter (SIRPOL) [46], brought the first performance of wide-field imaging linear and circular polarimetry in near-infrared (NIR). The IRSF telescope is located at the South African Astronomical Observatory. The observations by IRSF/SIRPOL are reviewed in section 4.

In the star-forming region, the cloud harbors the forming young star or stars. When a cavity exists around the forming star, the light from the forming star preferably propagates in the cavity. The scattered light at the wall of the cavity will be linearly polarized. In addition, the cavity decreases opacity in the line of sight. Therefore, when we observe the star-forming region from the outside of the region, the observed LP distribution indicates the existence of a cavity in the cloud [44]. In fact, on star formation, a bipolar cavity around the forming star is often produced in a parent cloud, by outflows and jets from the forming star. As numerical simulations show [47], the LP region can appear depending on inclination of the system. Further, the direction of LP tends to be centrosymmetric around the major light source [44]. The star-forming region often has many young stars. The direction map of observed LP indicates the dominant light source of the observed region.

Although discussion of many proposed mechanisms leading to homochirality of the terrestrial life is useful for consideration of the terrestrial life on our Earth and various extrasolar life on ESPs around nearby stars which would be directly observed in future [48], in this chapter, we focus on the idea of the extraterrestrial origin leading to homochirality, considering circular polarization in space. It would be important to consider a consistent theory with the birth and evolution of Earth in the solar system. In section 2, the organic materials and water in space is reviewed, as possible necessary material for life and chirality. In section 3, enantiomeric excess (EE) in meteorites and asymmetric photochemistry by circularly polarized light (CPL), and amplification of EE are reviewed. In section 4, the source of CPL in space is reviewed. These are summarized in section 5.

## 2. ORGANIC MATERIALS AND WATER IN SPACE

### 2.1 Extraterrestrial Organic Matter, Amino Acid, Amino Acid Precursor, and Sugar

Amino acids or amino acid precursors, molecules that provide amino acids after acid hydrolysis, are thought to exist in space [49]. Organic matters seem to be popular in space, considering detections in our solar and interstellar materials [50, 49].

In our solar system, the carbonaceous chondrite meteorites, which are thought to be the most ancient meteorites indicating starting materials of the presolar molecular cloud [51, 52], present organic matters including amino acids [53, 54, 55, 56, 57]. Interplanetary dust particles, which are thought to be samples of primitive objects [58, 59], also show organics [60, 61, 62, 63]. The recent returned samples by spacecraft from comet 81P/Wild 2, which can be the accreted materials during the formation of our solar system, showed organics [64, 65, 66, 67] including amino acid [68], glycine confirmed by the stable carbon isotopic ratios [69].

Glycine ($NH_2CH_2COOH$) is the simplest and achiral (i.e., not chiral) amino acid, which is used in life. Since 1979 [70], glycine has been investigated in interstellar medium involving controversy [71, 49]. The detection of interstellar glycine was proposed in several interstellar clouds including the Orion KL [72], followed by different perspectives [73].

Amino acetonitrile ($NH_2CH_2CN$), which is a possibly direct amino acid precursor of glycine, was detected in one of Galactic major center sources of activity, Sagittarius B2(N), indicating of formation by grain surface chemistry [74]; Sagittarius B2 is a very massive (several million solar masses) and extremely active region of massive-star formation near the Galactic center (projected distance of about 100 pc from the Galactic center).

Organic molecules have been detected in interstellar and circumstellar medium [49], for example, in high mass star forming region such as the Orion KL [75, 76], low mass star forming region such as IRAS 16293-2422 [77, 78], and protoplanetary disks [79, 80].

Interstellar glycolaldehyde ($CH_2OHCHO$), the simplest possible aldeyde sugar, was detected in the Sagittarius B2(N) [81]. Toward the Sagittarius B2(N-LMH), interstellar ethylene glycol ($HOCH_2CH_2OH$), the sugar alcohol of glycolaldehyde, was detected [82]; The three-carbon keto ring, cyclopropenone ($c-H_2C_3O$) was detected, with no detection of the three-carbon sugar glyceraldehyde [83].

**2.2 Extraterrestrial Water**

Water, which could contribute to development of chirality and life, seems to be broadly distributed in the universe [84], apart from its phase. In our solar system, our Earth has solid, liquid, and vapor water. High resolution images of Martian surface by the Mars Global Surveyor suggested the liquid water seepage and surface runoff on Mars [85]. The ultraviolet imaging spectrograph of the Cassini space craft revealed a water vapor plume in the south polar region of Saturn's moon, Enceladus [86]. The observation of the disk-averaged light of another moon of Saturn, Titan, indicated water icy bedrock [87]. The solid water ice deposits were directly detected on the surface of comet 9p/Tempel 1 on the Deep Impact mission, suggesting that the surface deposits are loose aggregates [88]. In our outer solar system, the presence of crystalline water ice was reported on the Kuiper belt object (50000) Quaoar in the Kuiper belt, which is consist of solid bodies beyond Neptune and yields comets [89].

In outside of our solar system, the water vapor was detected in the atmosphere of an ESP, a transiting hot Jupiter [90]. Water vapor and organic molecules were indicated in the inner protoplanetary disk around a classical T Tauri star, which is thought to be a young sun-like star, using the Spitzer Space Telescope [91]. The water ice grains were detected in the circumstellar disk around a Herbig Ae star, which is thought to be a young massive star, using Subaru Telescope [92]. $H_2O$ maser emission is detected in massive star-forming regions [93] including the Orion KL [94]. The ortho-$H_2O$ emission was detected in molecular cloud cores [95]. Numerical simulations of terrestrial planet formation indicate the possibility of broadly distributed water in planetary system, depending on conditions [96].

# 3. ENANTIOMERIC EXCESS AND ASYMMETRIC PHOTOCHEMISTRY

## 3.1 Enantiomeric Excess in Meteorites and the Late Heavy Bombardment

Detections of enantiomeric excesses (EEs) of amino acids in several meteorites (Murchison, Murray, Orgueil) have been reported, with small EEs of the same handedness as terrestrial life [97, 98, 99, 100, 101, 56]. The detections of EEs from meteorites which have fallen down to Earth indicate that EEs in meteorites survive through infall to Earth's atmosphere, and EEs in meteorites survive through long time after EEs were formed, as the stability of chiral amino acids against radionuclides decay in comets and asteroids in 4.6 billion years is investigated by experiments [102]. The detections of EEs support the hypothesis of the extraterrestrial origin of life which was seeded by delivery of organics from outer space.

If meteorites (and small objects in our solar system) brought seeds of life to Earth, the efficiency and the period is to be considered. The first terrestrial life appeared on Earth probably before about 3.8 billion years ago, indicated by sedimentary protolith [103, 104] and sedimentary rocks [105] from the Isua supracrustal belt in west Greenland. About 3.9 billion years ago, a lot of meteorites fell down on Earth's moon, and then Earth [106], as also indicated by sedimentary rocks from Isua greenstone belt in west Greenland [107]. This is called *the late heavy bombardment phase*, which would occur throughout the inner solar system [108]. In this late moment from the birth of Earth (about 4.6 billion years ago), many drops of meteorites with peculiar EEs could bring peculiar EEs over Earth before the emergence of life. Coincidentally, the ocean of Earth probably appeared before about 3.8 billion years ago [109]. It could be the period for the appearance of terrestrial life. The late heavy bombardment is thought to be brought due to the migration of giant planets [110, 111]. If so, the late heavy bombardment could be connected with evolution of some type of planetary systems and could occur in another planetary system.

## 3.2 Amino Acid and Amino Acid Precursor by Photochemistry

The extraterrestrial origin leading to homochirality requires the material and the mechanism of production of EEs of amino acids, *in space*. Regarding the material, as denoted in section 2, organic materials involving amino acids, amino acid precursors, or their elements can be available in space, apart from chirality.

Apart from EE, amino acids (or amino acid precursors) have been obtained by photochemistry in laboratory experiments for interstellar chemistry. In 2002, the analogues of icy interstellar grains (an ice film consisting of amorphous $H_2O$, $NH_3$, $CH_3OH$, and HCN) at 15 K were irradiated by ultraviolet (UV) light in vacuum. After warming the ices to room temperature, and after hydrolysis, racemic amino acids (glycine, alanine, and serine) were obtained [112]. In another experiment, an interstellar ice analogue (an ice mixture containing $H_2O$, $CH_3OH$, $NH_3$, CO, and $CO_2$) was irradiated by UV light at 12 K in vacuum. After warming the system to room temperature, 16 amino acids were identified [113]. Only after acid hydrolysis, amino acids were detected at considerable amounts. Results in these two experiments were also confirmed in experiments in 2007 [114]. In 2005, an interstellar ice analogue containing $H_2O$, $CH_3OH$, $NH_3$, CO, $CO_2$ were irradiated by UV light, at 12 K in vacuum. After warming up to room temperature, N-heterocyclic molecules and amines were detected in water extracts [115]. Carboxylic acid salts as part of the refractory products were shown using infrared spectroscopy in 2003 [116].

In 2007, an ice mixture containing $H_2O$, $CO_2$, and $NH_3$ was irradiated with UV light at 16 K in vacuum. This starting ice mixture did not contain any organic compound such as methanol ($CH_3OH$) and methane ($CH_4$). After 6 times repeat of warming up to room temperature, cooling down, and irradiation, finally, the proteinaceous amino acids were identified in the production [117]. A detailed analysis of amino acids which was produced by the UV irradiation of interstellar ice analogues was reported in 2008, highlighting the contribution of acid hydrolysis to yield amino acids [118]. The experiment using naphthalene ($C_{10}H_8$), the smallest polycyclic aromatic hydrocarbon (so called, PAH), was also performed with UV in vacuum at 15 K, producing amino acids. The naphthalene was mixed in an ice mixture of $H_2O$ and $NH_3$ [119].

The mechanism for the formation of the amino acids glycine, serine, and alanine in interstellar ice analogs was investigated using isotopic labeling techniques in 2007, indicating the multiple pathways to amino acid formation [120]. In 2009, the structures of the products and their formation pathways were investigated using deuterium-labeling experiments, indicating the initial photochemical cleavages of C-H and N-H bonds, for glycine [121]. The efficiency of photochemical synthesis of glycine on the ice surfaces and steady-state equilibrium between photosynthesis and photodestruction of glycine were pointed out.

Photostability of amino acids against UV photodestruction were investigated [122], as well as other small biomolecules were investigated [123]. Amino acids may be preferably destructed in UV, so some protections [122] or escapes during accretion [124] may be necessary. On the other hand, aminoacetonitrile ($H_2NCH_2CN$), which is an amino acid precursor to the amino acid glycine, is more stable than amino acid against UV photolysis [125].

## 3.3 Enantiomeric Excess by Asymmetric Photochemistry

The laboratory experiments show that EE can be yielded by asymmetric photochemistry using CPL [126, 127]: asymmetric photolysis, asymmetric synthesis, and asymmetric photoisomerization.

In asymmetric photolysis [18, 19, 128], one of the enantiomer is preferentially destructed under CPL, and then the other is enriched. The handedness of EE depends on the handedness of the CPL. Even elliptically-polarized light can induce asymmetric photolysis with a lesser degree than CPL [129]. In 2005, the amino acid leucine in the solid state was photolysed by UV CPL in vacuum, while other experiments were often performed in solutions [130]. The experiment showed the highest gain of ~2.6% in D-leucine.

Remarkably, several recent laboratory experiments have been performed with the interstellar analogues, although the interpretation of the results for effective mechanisms is more complex than simple experiments. In 2006, the interstellar ice analogues (gas mixtures) were irradiated by UV CPL under interstellar-like conditions (80 K, ~$10^{-7}$ mbar). However, very small EEs (at most ~1%) were produced, comparable with the detection limit (~1%) of the chromatography used in the experiment (GC-MS) [131].

In 2007, the possibility of asymmetric synthesis of amino acid precursors in interstellar complex organic under CPL was demonstrated in laboratory [132]. In their experiment, initially, gas mixtures (at room temperature) of carbon monoxide, ammonia and water, which are identified in the interstellar medium, were irradiated with high energy protons, yielding complex organic compounds. A liquid portion of the proton-irradiated sample was irradiated with UV CPL. Following acid hydrolysis, alanine showed EEs of +0.44% and -0.65% by right-handed CPL and left-handed CPL, respectively. Comparing an unhydrolyzed fraction with a product following acid hydrolysis, they speculated that combined amino acid analogs, rather than free amino acids, were present in the CPL-irradiated samples.

**3.4 Asymmetric Amplification of Tiny Enantiomeric Excess for Amimo Acid and Sugar**

Even if the EE is small as described in the previous section 3.3, the EE can be amplified by some process: asymmetric amplification [133, 134]. Experiments have shown that low EEs can be amplified by asymmetric autocatalysis (autocatalysed reactions) [135, 136]. In this reaction, a chiral product serves as a chiral catalyst for its own formation in the reaction. Asymmetric autocatalysis of 2-(*tert*-butylethynyl)-5-pyrimidyl alkanol showed that low EE (~0.00005%) was amplified to large EE (>99.5%) [137].

The small EEs can be amplified into solutions, because the solubility of an exclusive enaniomer is higher than that of the racemic compound crystal [138, 139, 140, 141]. This solubility-based amplification (solid-liquid phase behavior) is asymmetric aldol reaction and can occur in aqueous systems. Small EE (~1%) of serine was amplified to large EE (>99%) under solid-liquid equilibrium conditions [138].

Homogeneity of right-handed sugars may be initiated by low EEs of amino acids as a catalyst [142, 143, 144, 145]. The solubility-based amplification for amino acids in water [138] can also work for nucleosides, which would lead to the RNA world [146]. These reactions support that exogenous injection of low EEs of amino acids on early Earth yields the homochirality of amino acid and sugar, and then the terrestrial life.

## 4. CIRCULARLY POLARIZED LIGHT SOURCE IN SPACE

**4.1 Possible Sources of Circular Polarization**

Neutron stars and magnetic white dwarfs seems not to be a CP source yielding of EEs for early solar system, considering no detection of CP in optical or few encounters with a molecular cloud or star-forming region [147, 124]. On the other hand, CP of young stellar objects (YSOs), which are thought to be young stars, has been detected by circular polarimetry. In table 1, the degree of CP of YSOs is summarized from previous circular polarimetry. In September 2010, the *imaging* circular polarimetry of YSOs is still scarce, although hundreds of *point-like sources* in the core of the Orion nebula were performed in [148]. According to the previous observations, more massive YSOs appear to have larger CP. The possibility of the contribution of the stronger magnetic field with the formation of higher-mass stars was pointed out, which leads to more efficient alignment of dust grains [124, 149]. The

Orion nebula is one of YSOs which have the largest CP.

.       Previous imaging circular polarimetry of YSOs (in Table 1) and numerical simulations producing CP in YSOs [159, 160, 47, 161, 162, 149] indicate that a YSO will not usually produce a net CP because it will have regions of right-handed and left-handed CP that cancel globally. Such patchy distribution of right-handed and left-handed CP could yield patchy distribution of right-handed and left-handed EEs, so that meteorites in late heavy bombardment could yield inefficient EEs. To inject efficient EEs on early Earth, the entire irradiation on early solar system or its materials by one-handed CP would be necessary.

**Table 1. The degree of CP (%) of young stellar objects in previous observations.**

| Mass | Object name | The degree of CP (*color band*) | Reference |
| --- | --- | --- | --- |
| **high** | OMC-1 in Orion | 17($K_n$), 5($H$), 2($J$) | [147], [150], [148] |
|  | NGC 6334-V | 23($K$) | [151] |
| **intermediate** | HH 135-136 | 8 ($K_n$), 3 ($H$), 2.5 ($J$) | [149] |
|  | R CrA | 5 ($H$) | [152] |
| **low** | GSS 30 | 1.7 ($K_n$), 0.8 ($H$) | [153] |
|  | Cha IRN | 1 ($H$) | [154] |
| high | R Mon | 0.4 ($R$) | [155] |
| low | HL Tau | no detection, <0.5 ($J$, $H$, $K$) | [156] |
| intermediate | PV Cep | < 1 ($V$, $I$) | [157] |
|  | V633 Cas | < 1 ($V$, $I$) | [157] |
| low/intermediate | L1551 IR5 | < 3 ($V$, $I$) | [157] |
| high | GL 2591 | < 1 ($I$) | [157] |
| low | ~hundreds point-like sources in Orion | <~1.5% ($H$, $K_s$) | [148] |

The table 1 in [157] is updated in September 2010 (see also [158]). The nebulae in the first three columns (denoted by bold face) were detected by imaging polarimetry.

## 4.2 Wide Extension of Linear and Circular Polarization in the Orion Nebula

LP images of the Orion nebula in NIR using IRSF/SIRPOL revealed the large LP region, which extends over about 0.7 pc, in 2006 [163]. The three color bands were used: *J* (1.25μm), *H* (1.63μm), $K_s$ (2.14μm). The nebulae emitting LP are located around the massive star-forming region, the BN/KL region. This extension of LP indicates the existence of large cavities around the young massive stars in the BN/KL region. Moreover, the observation showed the linearly polarized Orion bar, several small linearly polarized nebulae, and the low LP near the Trapezium. The Trapezium is a group of massive young stars and is located near the BN/KL region.

CP images of the Orion nebula in $K_s$ band using IRSF/SIRPOL revealed the large CP region, which extends over about 0.4 pc, in 2010 (see Figure 1 in [148]). The observed CP extends over a region about 400 times the size of the solar system, when the size of the solar system is assumed to be about 200 AU in diameter. This extension of the CP region is almost comparable to that of the LP region. This CP region is located around the BN/KL region. The degrees of CP range from +17% to -5%. The linearly polarized Orion bar in linear polarimetry shows no significant CP in circular polarimetry. The small linearly polarized nebulae in linear polarimetry show no significant CP. The aperture circular polarimetry of the 353 point-like sources, many of which are low-mass young stars, showed that the degree of CP for each source is generally small, less than ~1.5% (see Figure 2 in [148]).

Although the CP distribution of low mass young stars would not be spatially resolved in aperture polarimetry, the result in [148] showed that the point-like sources do not have generally large degree of CP. In other words, the point-like source does not emit large one-handed CP from its entire face to us. Even if low mass young stars have inherent CP locally, the CP degree would be low as spatially resolved nebulae showed low CP degree in previous observations (see Table 1), or/and the distribution of right-handed and left-handed CP would be patchy, in which the right-handed and left-handed CP are cancelled leading to the appearance of low CP degree when the distribution is not spatially resolved. Such CP distribution would yield inefficient EEs on early Earth, as described in section 4.1.

The major contribution to produce the observed CP in the Orion nebula is thought to be dichroic extinction [164, 165]. The light from the central star/stars propagates inside the cavity around the central star/stars. The light is scattered at the wall of the cavity, linearly polarized, and goes though the surrounding cloud. When the dust grains are non-spherical and somewhat aligned in the cloud, the incident linearly polarized light is (partially) circularly polarized. In other words, the aligned dust grains in the cloud behave like 1/4 wave plate for the incident LP. This situation can be connected with a simple relation between LP, CP and the color excess in the imaging observation. In fact, the observed correlation between LP, CP and the color excess agrees with the relation [165].

UV light in star-forming regions can not be directly observed. The dust grains drifting in star-forming regions prevent us to observe by UV light. Therefore, observations of star-forming regions are often performed in near-infrared. Numerical simulations to produce CP in a modeled space are helpful to investigate UV CP in star-forming regions [147, 162].

## 5. CONCLUSION

Detections of EEs in several meteorites support the possibility that the seed of life was injected from space. CPL could bring the EE of prebiotic molecules in space, as shown in experimental works on photochemistry under UV CPL. EE by photochemistry under CPL would be small. However, several mechanisms for amplification of the EE into almost pure enantiomers have been shown in experiments.

Recent astronomical observations by imaging circular polarimetry of star-forming regions are now revealing the distribution of CPL in space. The observed significant CP in the core of the Orion nebula extends over a region about 400 times the size of the solar system. If a solar system would be formed in such nebula and be irradiated by CP, EE would be produced by asymmetric photochemistry and yield EE with meteorites and small objects onto an early planet/moon. This could result in life with homochirality.

The observed CP in the Orion nebula showed wide extension of both regions of right-handed and left-handed CP. Since the EE yielded by CP is dependent on the handedness of CP, this result might indicate that the handedness of possible EE produced by CP is different among those CP regions in the Orion nebula.

For deep insights on terrestrial life, extrasolar life, and the origin of life in the universe, it would be important to consider both of the place where life is able to live, *the habitable zone*, and the place where life is able to originate, what we shall call *the originable zone*. A habitable zone and an originable zone would not be equal. However, they would have partial overlap.

The Orion nebula seems to be located in the galactic habitable zone. The Orion nebula harbors a lot of stellar systems, some of which would harbor the circumstellar habitable zone. When the CPL in the massive star forming region, the BN/KL region, contributes to the EEs and leads to the homochirality for the terrestrial life, the Orion nebula would harbors the originatable zone for the terrestrial life. The Orion nebula, a young star-forming region, whose age is about one million years, seems to bring the extraterrestrial life in future. In billions of years, some extraterrestrial intelligences on some ESPs from the Orion nebula might watch the end of our Earth and us, who might watch the birth of them.

## ACKNOWLEDGMENTS

The author thanks Jun Fukue for useful comments and encouragements.

## REFERENCES

1. Mayor, M.; Queloz, D. *Nature* 1995, 378, 355-359.

2. Charbonneau, D.; Brown, T. M.; Latham, D. W.; Mayor, M. *Astrophys J* 2000, 529, 45-48.

3. Kalas, P.; Graham, J. R.; Chiang, E.; Fitzgerald, M. P.; Clampin, M.; Kite, E. S.; Stapelfeldt, K.; Marois, C.; Krist, J. *Science* 2008, 322, 1345-1348.

4. Arnold, L. *Space Sci Rev* 2008, 135, 323-333.

5. Kaltenegger, L.; Traub, W. A.; Jucks, K. W. *Astrophys J*, 2007, 658, 598-616.

6. Gonzalez, G. *Orig Life Evol Biosph* 2005, 35, 555-606.

7. Javaux, E. J.; Dehant, V. *Astron Astrophys Rev*, 2010, 18, 383-416.

8. Kasting, James F.; Whitmire, Daniel P.; Reynolds, Ray T. *Icarus*, 1993, 101, 108-128.

9. Gonzalez, G.; Brownlee, D.; Ward, P. *Icarus*, 2001, 152, 185-200.

10. Kasting, J. F.; Catling, D. *Annu Rev Astron Astrophys*, 2003, 41, 429-463.

11. Kaltenegger, L. *Astrophys J L*, 2010, 712, L125-L130.

12. Buccino, A. P.; Lemarchand, G. A.; Mauas, P. J. D. *Icarus*, 2007, 192, 582-587.

13. Kaltenegger, L.; Eiroa, C.; Ribas, I.; Paresce, F.; Leitzinger, M.; Odert, P.; Hanslmeier, A.; Fridlund, M.; Lammer, H.; Beichman, C.; Danchi, W.; Henning, T.; Herbst, T.; Léger, A.; Liseau, R.; Lunine, J.; Penny, A.; Quirrenbach, A.; Röttgering, H.; Selsis, F.; Schneider, J.; Stam, D.; Tinetti, G.; White, G. J. *Astrobiology* 2010, 10, 103-112.

14. Fujii, N. *Orig Life Evol Biosph* 2002, 32, 103-127.

15. Fujii, N.; Saito, T. *The Chemical Record* 2004, 4, 267-278.

16. Fujii, N. *Biol Pharm Bull* 2005, 28, 1585-1589.

17. Fuchs, S. A.; Berger, R.; Klomp, L. W. J.; de Koning, T. J. *Mol Genet Metab* 2005, 85, 168-180.

18. Bonner, W. A. *Orig Life Evol Biosph* 1991, 21, 59-111.

19. Bonner, W. A. *Orig Life Evol Biosph* 1995, 25, 175-190.

20. Meierhenrich, U. J.; Thiemann, W. H.-P. *Orig Life Evol Biosph* 2004, 34, 111-121.

21. Barron, L. D. *Space Sci Rev* 2008, 135, 187-201.

22. Bouvier, A.; Wadhwa, M. *Nature Geoscience* 2010, 3, 637-641.

23. Morishima, R.; Stadel, J.; Moore, B. *Icarus*, 2010, 207, 517-535.

24. Moro-Martín, A.; Malhotra, R.; Bryden, G.; Rieke, G. H.; Su, K. Y. L.; Beichman, C. A.; Lawler, S. M. *Astrophys J* 2010, 717, 1123.

25. Matsui, T.; Abe, Y. *Nature*, 1986, 319, 303-305.

26. Abe, Y. *Physics of The Earth and Planetary Interiors*, 1997, 100, 27-39.

27. Greenwood, R. C.; Franchi, I. A.; Jambon, A.; Buchanan, P. C. *Nature*, 2005, 435, 916-918.

28. Wetherill, G. W. *Science*, 1985, 228, 877-879.

29. Kokubo, E.; Ida, S. *Icarus*, 2000, 143, 15-27.

30. Canup, R. M.; Asphaug, E. *Nature*, 2001, 412, 708-712.

31. Canup, R. M. *Annu Rev Astron Astrophys*, 2004, 42, 441-475.

32. Melosh, H. J. In Origin of the Earth; Newsom, H. E.; Jones, J. H.; Ed.; Oxford University Press: NY, 1990, 69-83.

33. Cameron, A. G. W. *Icarus*, 1997, 126, 126-137.

34. Wood, B. J.; Halliday, A. N. *Nature*, 2005, 437, 1345-1348.

35. Wood, B. J.; Walter, M. J.; Wade, J. *Nature*, 2006, 441, 825-833.

36. Oberbeck, V. R.; Fogleman, G. *Orig Life Evol Biosph* 1989, 19, 549-560.

37. Hester, J. J.; Desch, S. J. In: Krot, A. N. et al (ed) Chondrites and the protoplanetary disk, ASP, San Francisco, 2005, 341, 107-130.

38. McKee, C. F.; Ostriker, E. C. *Annual Review of Astronomy & Astrophysics*, 2007, 45, 565-687.

39. Tachibana, S.; Huss, G. R. *Astrophys J*, 2003, 588, 41-44.

40. Mostefaoui, S.; Lugmair, G. W.; Hoppe, P. *Astrophys J*, 2005, 625, 271-277.

41. Tachibana, S.; Huss, G. R.; Kita, N. T.; Shimoda, G.; Morishita, Y. *Astrophys J*, 2006, 639, 87-90.

42. Dauphas, N.; Cook, D. L.; Sacarabany, A.; Fröhlich, C.; Davis, A. M.; Wadhwa, M.; Pourmand, A.; Rauscher, T.; Gallino, R. *Astrophys J*, 2008, 686, 560-569.

43. Hough, J. *Astronomy & Geophysics*, 2006, 47, 31-35.

44. Hough, J. H. *Journal of Quantitative Spectroscopy and Radiative Transfer*, 2007, 106, 122-132.

45. Nagayama, T.; Nagashima, C.; Nakajima, Y.; Nagata, T.; Sato, S.; Nakaya, H.; Yamamuro, T.; Sugitani, K.; Tamura, M. *Proceedings of the SPIE*, 2003, 4841, 459-464.

46. Kandori, R.; Kusakabe, N.; Tamura, M.; Nakajima, Y.; Nagayama, T.; Nagashima, C.; Hashimoto, J.; Hough, J.; Sato, S.; Nagata, T.; Ishihara, A.; Lucas, P.; Fukagawa, M. *Proceedings of the SPIE*, 2006, 6269, 159.

47. Whitney, B. A.; Wolff, M. J. *Astrophys J*, 2002, 574, 205-231.

48. Kaltenegger, L.; Eiroa, C.; Fridlund, C. V. M. *Astrophysics and Space Science*, 2010, 326, 233-247.

49. Herbst, E.; van Dishoeck, E. F. *Annu Rev Astron Astrophys*, 2009, 47, 427-480.

50. Ehrenfreund, P.; Charnley, S. B. *Annu Rev Astron Astrophys*, 2000, 38, 427-483.

51. Busemann, H.; Young, A. F.; O'D. Alexander, C. M.; Hoppe, P.; Mukhopadhyay, S.; Nittler, L. R. *Science*, 2006, 312, 727-730.

52. Nakamura-Messenger, K.; Messenger, S.; Keller, L. P.; Clemett, S. J.; Zolensky, M. E. *Science*, 2006, 314, 1439-1442.

53. Kvenvolden, K.; Lawless, J.; Pering, K.; Peterson, E.; Flores, J.; Ponnamperuma, C. *Nature*, 1970, 228, 923-926.

54. Botta, O.; Bada, J. L. *Surveys in Geophysics*, 2002, 23, 411-467.

55. Cody, G. D.; Alexander, C. M. O.; Tera, F. *Geochim Cosmochim Acta*, 2002, 66, 1851-1865.

56. Sephton, M. A. *Nat Prod Rep*, 2002, 19, 292-311.

57. Schmitt-Kopplina, P.; Gabelicab, Z.; Gougeonc, R. D.; Feketea, A.; Kanawatia, B.; Harira, M.; Gebefuegia, I.; Eckeld, G.; Hertkorna, N. *PNAS*, 2010, 107, 2763-2768.

58. Messenger, S. *Nature*, 2000, 404, 968-971.

59. Duprat, J.; Dobrică, E.; Engrand, C.; Aléon, J.; Marrocchi, Y.; Mostefaoui, S.; Meibom, A.; Leroux, H.; Rouzaud, J.-N.; Gounelle, M.; Robert, F. *Science*, 2010, 328, 742-745.

60. Maurette, M.; Duprat, J.; Engrand, C.; Gounelle, M.; Kurat, G.; Matrajt, G.; Toppani, A. *Planetary and Space Science*, 2000, 48, 1117-1137.

61. Flynn, G. J.; Keller, L. P.; Feser, M.; Wirick, S.; Jacobsen, C. *Geochim Cosmochim Acta*, 2003, 67, 4791-4806.

62. Messenger, S.; Stadermann, F. J.; Floss, C.; Nittler, L. R.; Mukhopadhyay, S. *Space Sci Rev*, 2003, 106, 155-172.

63. Floss, C.; Stadermann, F. J.; Bradley, J. P.; Dai, Z. R.; Bajt, S.; Graham, G.; Lea, A. S. *Geochim Cosmochim Acta*, 2006, 70, 2371-2399.

64. Sandford, S. A.; Aléon, J.; Alexander, C. M. O.'D.; Araki, T.; Bajt, S.; Baratta, G. A.; Borg, J.; Bradley, J. P.; Brownlee, D. E.; Brucato, J. R.; Burchell, M. J.; Busemann, H.; Butterworth, A.;

Clemett, S. J.; Cody, G.; Colangeli, L.; Cooper, G.; D'Hendecourt, L.; Djouadi, Z.; Dworkin, J. P.; Ferrini, G.; Fleckenstein, H.; Flynn, G. J.; Franchi, I. A.; Fries, M.; Gilles, M. K.; Glavin, D. P.; Gounelle, M.; Grossemy, F.; Jacobsen, C.; Keller, L. P.; Kilcoyne, A. L. D.; Leitner, J.; Matrajt, G.; Meibom, A.; Mennella, V.; Mostefaoui, S.; Nittler, L. R.; Palumbo, M. E.; Papanastassiou, D. A.; Robert, F.; Rotundi, A.; Snead, C. J.; Spencer, M. K.; Stadermann, F. J.; Steele, A.; Stephan, T.; Tsou, P.; Tyliszczak, T.; Westphal, A. J.; Wirick, S.; Wopenka, B.; Yabuta, H.; Zare, R. N.; Zolensky, M. E. *Science*, 2006, 314, 1720-1724.

65. Cody, G. D.; Ade, H.; O'D. Alexander, C. M.; Araki, T.; Butterworth, A.; Fleckenstein, H.; Flynn, G.; Gilles, M. K.; Jacobsen, C.; Kilcoyne, A. L. D.; Messenger, K.; Sandford, S. A.; Tyliszczak, T.; Westphal, A. J.; Wirick, S.; Yabuta, H. *Meteoritics & Planetary Science*, 2008, 43, 353-365.

66. Wirick, S.; Flynn, G. J.; Keller, L. P.; Nakamura-Messenger, K.; Peltzer, C.; Jacobsen, C.; Sandford, S. A.; Zolensky, M. E. *Meteoritics & Planetary Science*, 2009, 44, 1611-1626.

67. De Gregorio, B. T.; Stroud, R. M.; Nittler, L. R.; Alexander, C. M. O'D.; Kilcoyne, A. L. D.; Zega, T. J. *Geochim Cosmochim Acta*, 2010, 74, 4454-4470.

68. Glavin, D. P.; Dworkin, J. P.; Sandford, S. A. *Meteoritics & Planetary Science*, 2008, 43, 399-413.

69. Elsila, J. E.; Glavin, D. P.; Dworkin, J. P. *Meteoritics & Planetary Science*, 2009, 44, 1323-1330.

70. Brown, R. D.; Godfrey, P. D.; Storey, J. W. V.; Bassez, M.-P.; Robinson, B. J.; Batchelor, R. A.; McCulloch, M. G.; Rydbeck, O. E. H.; Hjalmarson, A. G. *Mon Not R Astron Soc*, 1979, 186, 5-8.

71. Snyder, L. E. *Orig Life Evol Biosph*, 1997, 27, 115-133.

72. Kuan, Y.-J.; Charnley, S. B.; Huang, H.-C.; Tseng, W.-L.; Kisiel, Z. *Astrophys J*, 2003, 593, 848-867.

73. Snyder, L. E.; Lovas, F. J.; Hollis, J. M.; Friedel, D. N.; Jewell, P. R.; Remijan, A.; Ilyushin, V. V.; Alekseev, E. A.; Dyubko, S. F. *Astrophys J*, 2005, 619, 914-930.

74. Belloche, A.; Menten, K. M.; Comito, C.; Müller, H. S. P.; Schilke, P.; Ott, J.; Thorwirth, S.; Hieret, C. *Astron Astrophys*, 2008, 482, 179-196.

75. Sutton, E. C.; Peng, R.; Danchi, W. C.; Jaminet, P. A.; Sandell, G.; Russell, A. P. G. *Astrophys J Suppl Ser*, 1995, 97, 455-496.

76. Schilke, P.; Groesbeck, T. D.; Blake, G. A.; Phillips, T. G. *Astrophys J Suppl*, 1997, 108, 301-337.

77. van Dishoeck, E. F.; Blake, G. A.; Jansen, D. J.; Groesbeck, T. D. *Astrophys J*, 1995, 447, 760-782.

78. Cazaux, S.; Tielens, A. G. G. M.; Ceccarelli, C.; Castets, A.; Wakelam, V.; Caux, E.; Parise, B.; Teyssier, D. *Astrophys J*, 2003, 593, 51-55.

79. Thi, W.-F.; van Zadelhoff, G.-J.; van Dishoeck, E. F. *Astron Astrophys*, 2004, 425, 955-972.

80. Öberg, K. I.; Qi, C.; Fogel, J. K. J.; Bergin, E. A.; Andrews, S. M.; Espaillat, C.; van Kempen, T. A.; Wilner, D. J.; Pascucci, I. *Astrophys J*, 2010, 720, 480-493.

81. Hollis, J. M.; Jewell, P. R.; Lovas, F. J.; Remijan, A. *Astrophys J*, 2004, 613, 45-48.

82. Hollis, J. M.; Lovas, F. J.; Jewell, P. R.; Coudert, L. H. *Astrophys J*, 2002, 571, 59-62.

83. Hollis, J. M.; Remijan, Anthony J.; Jewell, P. R.; Lovas, F. J. *Astrophys J*, 2006, 642, 933-939.

84. Encrenaz, T. *Annu Rev Astron Astrophys*, 2008, 46, 57-87.

85. Malin, M. C.; Edgett, K. S. *Science*, 2000, 288, 2330-2335.

86. Hansen, C. J.; Esposito, L.; Stewart, A. I. F.; Colwell, J.; Hendrix, A.; Pryor, W.; Shemansky, D.; West, R. *Science*, 2006, 311, 1422-1425.

87. Griffith, C. A.; Owen, T.; Geballe, T. R.; Rayner, J.; Rannou, P. *Science*, 2003, 300, 628-630.

88. Sunshine, J. M.; A'Hearn, M. F.; Groussin, O.; Li, J.-Y.; Belton, M. J. S.; Delamere, W. A.; Kissel, J.; Klaasen, K. P.; McFadden, L. A.; Meech, K. J.; Melosh, H. J.; Schultz, P. H.; Thomas, P. C.; Veverka, J.; Yeomans, D. K.; Busko, I. C.; Desnoyer, M.; Farnham, T. L.; Feaga, L. M.; Hampton, D. L.; Lindler, D. J.; Lisse, C. M.; Wellnitz, D. D. *Science*, 2006, 311, 1453-1455.

89. Jewitt, D. C.; Luu, J. *Nature*, 2004, 432, 731-733.

90. Tinetti, G.; Vidal-Madjar, A.; Liang, M.-C.; Beaulieu, J.-P.; Yung, Y.; Carey, S.; Barber, R. J.; Tennyson, J.; Ribas, I.; Allard, N.; Ballester, G. E.; Sing, D. K.; Selsis, F. *Nature*, 2007, 448, 169-171.

91. Carr, J. S.; Najita, J. R. *Science*, 2008, 319, 1504-1506.

92. Honda, M.; Inoue, A. K.; Fukagawa, M.; Oka, A.; Nakamoto, T.; Ishii, M.; Terada, H.; Takato, N.;

Kawakita, H.; Okamoto, Y. K.; Shibai, H.; Tamura, M.; Kudo, T.; Itoh, Y. *Astrophys J L*, 2009, 690, 110-113.

93. Beuther, H.; Walsh, A.; Schilke, P.; Sridharan, T. K.; Menten, K. M.; Wyrowski, F. *Astron Astrophys*, 2002, 390, 289-298.

94. Wright, M. C. H.; Plambeck, R. L.; Wilner, D. J. *Astrophys J,* 1996, 469, 216-237.

95. Snell, R. L.; Howe, J. E.; Ashby, M. L. N.; Bergin, E. A.; Chin, G.; Erickson, N. R.; Goldsmith, P. F.; Harwit, M.; Kleiner, S. C.; Koch, D. G.; Neufeld, D. A.; Patten, B. M.; Plume, R.; Schieder, R.; Stauffer, J. R.; Tolls, V.; Wang, Z.; Winnewisser, G.; Zhang, Y. F.; Melnick, G. J. *Astrophys J*, 2000, 539, 101-105.

96. Raymond, S. N.; Quinn, T.; Lunine, J. I. *Icarus*, 2004, 168, 1-17.

97. Cronin, J. R.; Pizzarello, S. *Science*, 1997, 275, 951-955.

98. Pizzarello, S.; Cronin, J. R. *Geochim Cosmochim Acta*, 2000, 64, 329-338.

99. Pizzarello, S.; Zolensky, M.; Turk, K. A. *Geochim Cosmochim Acta*, 2003, 67, 1589-1595.

100. Pizzarello, S.; Huang, Y.; Alexandre, M. R. *PNAS*, 2008, 105, 3700-3704.

101. Glavin, D. P.; Dworkin, J. P. *Proc Natl Acad Sci USA*, 2009, 106, 5487-5492.

102. Iglesias-Groth, S.; Cataldo, F.; Ursini, O.; Manchado, A. 2010, arXiv:1007.4529

103. Mojzsis, S. J.; Arrhenius, G.; McKeegan, K. D.; Harrison, T. M.; Nutman, A. P.; Friend, C. R. L. *Nature*, 1996, 384, 55-59.

104. Mojzsis, S. J.; Harrison, T. M. *Earth and Planetary Science Letters*, 2002, 202, 563-576.

105. Rosing, M. T. *Science*, 1999, 283, 674-676.

106. Cohen, B. A.; Swindle, T. D.; Kring, D. A. *Science*, 2000, 290, 1754-1756.

107. Schoenberg, R.; Kamber, B. S.; Collerson, K. D.; Moorbath, S. *Nature*, 2002, 418, 403-405.

108. Kring, D. A.; Cohen, B. A. *Journal of Geophysical Research*, 2002, 107, 4-1 - 4-6

109. Harald F.; Maarten de W.; Hubert S.; Minik R.; Karlis M. *Science*, 2007, 315, 1704-1707.

110. Gomes, R.; Levison, H. F.; Tsiganis, K.; Morbidelli, A. *Nature*, 2005, 435, 466-469.

111. Strom, R. G.; Malhotra, R.; Ito, T.; Yoshida, F.; Kring, D. A. *Science*, 2005, 309, 1847-1850.

112. Bernstein, M. P.; Dworkin, J. P.; Sandford, S. A.; Cooper, G. W.; Allamandola, L. J. *Nature*, 2002, 416, 401-403.

113. Muñoz Caro, G. M.; Meierhenrich, U. J.; Schutte, W. A.; Barbier, B.; Arcones Segovia, A.; Rosenbauer, H.; Thiemann, W. H.-P.; Brack, A.; Greenberg, J. M. *Nature*, 2002, 416, 403-406.

114. Nuevo, M.; Meierhenrich, U. J.; D'Hendecourt, L.; Muñoz Caro, G. M.; Dartois, E.; Deboffle, D.; Thiemann, W. H.-P.; Bredehöft, J.-H.; Nahon, L. *Advances in Space Research*, 2007, 39, 400-404.

115. Meierhenrich, U. J.; Muñoz Caro, G. M.; Schutte, W. A.; Thiemann, W. H.; Barbier, B.; Brack, A. *Chemistry*, 2005, 11, 4895-4900.

116. Muñoz Caro, G. M.; Schutte, W. A. *Astron Astrophys*, 2003, 412, 121-132.

117. Nuevo, M.; Chen, Y.-J.; Yih, T.-S.; Ip, W.-H.; Fung, H.-S.; Cheng, C.-Y.; Tsai, H.-R.; Wu, C.-Y. R. *Advances in Space Research*, 2007, 40, 1628-1633.

118. Nuevo, M.; Auger, G.; Blanot, D.; D'Hendecourt, L. *Origins of Life and Evolution of Biospheres*, 2008, 38, 37-56.

119. Chen, Y.-J.; Nuevo, M.; Yih, T.-S.; Ip, W.-H.; Fung, H.-S.; Cheng, C.-Y.; Tsai, H.-R.; Wu, C.-Y. R. *Mon Not R Astron Soc*, 2008, 384, 605-610.

120. Elsila, J. E.; Dworkin, J. P.; Bernstein, M. P.; Martin, M. P.; Sandford, S. A. *Astrophys J*, 2007, 660, 911-918.

121. Lee, C.-W.; Kim, J.-K.; Moon, E.-S.; Minh, Y. C.; Kang, H. *Astrophys J*, 2009, 697, 428-435.

122. Ehrenfreund, P.; Bernstein, M. P.; Dworkin, J. P.; Sandford, S. A.; Allamandola, L. J. *Astrophys J*, 2001, 550, 95-99.

123. Schwell, M.; Jochims, H.-W.; Baumgärtel, H.; Dulieu, F.; Leach, S. *Planetary and Space Science*, 2006, 54, 1073-1085.

124. Bailey, J. *Orig Life Evol Biosph*, 2001, 31, 167-183.

125. Bernstein, M. P.; Ashbourn, S. F. M.; Sandford, S. A.; Allamandola, L. J. *Astrophys J*, 2004, 601,

365-370.

126. Feringa, B.L.; Delden, R.A. van *Angew Chem Int Ed*, 1999, 38, 3418-3438.

127. Griesbeck, A. G.; Meierhenrich, U. J. *Angew Chem Int Ed*, 2002, 41, 3147-3154.

128. Cerf, C.; Jorissen, A. *Space Sci Rev*, 2000, 92, 603-612.

129. Bonner, W. A.; Bean, B. D. *Orig Life Evol Biosph*, 2000, 30, 513-517.

130. Meierhenrich, U. J.; Nahon, L.; Alcaraz, C.; Bredehöft, J. H.; Hoffmann, S. V.; Barbier, B.; Brack, A. *Angew Chem Int Ed*, 2005, 44, 5630-5634.

131. Nuevo, M.; Meierhenrich, U. J.; Muñoz Caro, G. M.; Dartois, E.; D'Hendecourt, L.; Deboffle, D.; Auger, G.; Blanot, D.; Bredehöft, J.-H.; Nahon, L. *Astron Astrophys*, 2006, 457, 741-751.

132. Takano, Y.; Takahashi, J.; Kaneko, T.; Marumo, K.; Kobayashi, K. *Earth and Planetary Science Letters*, 2007, 254, 106-114.

133. Melchiorre, P.; Marigo, M.; Carlone, A.; Bartoli, G. *Angew Chem Int Ed Engl,* 2008, 47, 6138-6171.

134. Soai, K.; Kawasaki, T. *Chirality*, 2006, 18, 469-478.

135. Soai, K.; Shibata, T.; Morioka, H.; Choji, K. *Nature*, 1995, 378, 767-768.

136. Shibata, T.; Yamamoto, J.; Matsumoto, N.; Yonekubo, S.; Osanai, S.; Soai, K. *J Am Chem Soc*, 1998, 120, 12157-12158.

137. Kawasaki, T.; Soai, K. *Journal of Fluorine Chemistry*, 2010, 131, 525-534.

138. Klussmann, M.; Iwamura, H.; Mathew, S. P.; Wells, D. H.; Pandya, U.; Armstrong, A.; Blackmond, D. G. *Nature*, 2006, 441, 621-623.

139. Breslow, R.; Levine, M. S. *PNAS*, 2006, 103, 12979-12980.

140. Dziedzica, P.; Zoua. W.; Ibrahema, I.; Sundéna, H.; Córdova, A. *Tetrahedron Letters*, 2006, 47, 6657-6661.

141. Aratake, S.; Itoh, T.; Okano, T.; Usui, T.; Shoji, M.; Hayashi, Y. *Chem Commun*, 2007, 2524-2526.

142. Weber, A. L. *Orig Life Evol Biosph* 2001, 31, 71-86.

143. Pizzarello, S.; Weber, A. L. *Science* 2004, 303, 1151.

144. Córdova, A.; Engqvist, M.; Ibrahem, I.; Casas, J.; Sundén, H. *Chem Commun* 2005, 2047–2049.

145. Córdova, A.; Zou, W.; Dziedzic, P.; Ibrahem, I.; Reyes, E.; Xu, Y. *Chem Eur J* 2006, 12, 5383–5397.

146. Breslow, R.; Cheng, Z.-L. *PNAS*, 2009, 106, 9144-9146.

147. Bailey, J.; Chrysostomou, A.; Hough, J. H.; Gledhill, T. M.; McCall, A.; Clark, S.; Menard, F.; Tamura, M. *Science*, 1998, 281, 672-674.

148. Fukue, T.; Tamura, M.; Kandori, R.; Kusakabe, N.; Hough, J. H.; Bailey, J.; Whittet, D. C. B.; Lucas, P. W.; Nakajima, Y.; Hashimoto, J. *Origins of Life and Evolution of Biospheres*, 2010, 40, 335-346.

149. Chrysostomou, A.; Lucas, P. W.; Hough, J. H. *Nature*, 2007, 450, 71-73.

150. Chrysostomou, A.; Gledhill, T. M.; Ménard, F.; Hough, J. H.; Tamura, M.; Bailey, J. *Mon Not R Astron Soc*, 2000, 312, 103-115.

151. Ménard, F.; Chrysostomou, A.; Gledhill, T.; Hough, J. H.; Bailey, J. In: Lemarchand G, Meech K (ed) Bioastronomy 99: a new era in the search for Life in the Universe, San Francisco, ASP Conf. 2000, 213, 355-358.

152. Clark, S.; McCall, A.; Chrysostomou, A.; Gledhill, T.; Yates, J.; Hough, J. *Mon Not R Astron Soc*, 2000, 319, 337-349.

153. Chrysostomou, A.; Menard, F.; Gledhill, T. M.; Clark, S.; Hough, J. H.; McCall, A.; Tamura, M. *Mon Not R Astron Soc*, 1997, 285, 750-758.

154. Gledhill, T. M.; Chrysostomou, A.; Hough, J. H. *Mon Not R Astron Soc*, 1996, 282, 1418-1436.

155. Menard, F.; Bastien, P.; Robert, C. *Astrophys J*, 1988, 335, 290-294.

156. Takami, M.; Gledhill, T.; Clark, S.; Mnard, F; Hough, J. H. in Star Formation 1999, Ed. T. Nakamoto, 1999, 205.

157. Clayton, G. C.; Whitney, B. A.; Wolff, M. J.; Smith, P.; Gordon, K. D. In: Adamson, A. et al. (ed) Astronomical polarimetry: current status and future directions. ASP, San Francisco, 2005, ASP Conf.

Ser. 343, 122-127.

158. Fukue, T. "Polarimetric Study of Star/Planet-Forming Regions", PhD Thesis, Kyoto University, 2009.

159. Fischer, O.; Henning, T.; Yorke, H. W. *Astron Astrophys*, 1996, 308, 863-885.

160. Wolf, S.; Voshchinnikov, N. V.; Henning, Th. *Astron Astrophys*, 2002, 385, 365-376.

161. Lucas, P. W.; Fukagawa, M.; Tamura, M.; Beckford, A. F.; Itoh, Y.; Murakawa, K.; Suto, H.; Hayashi, S. S.; Oasa, Y.; Naoi, T.; Doi, Y.; Ebizuka, N.; Kaifu, N. *Mon Not R Astron Soc*, 2004, 352, 1347-1364.

162. Lucas, P. W.; Hough, J. H.; Bailey, J.; Chrysostomou, A.; Gledhill, T. M.; McCall, A. *Origins of Life and Evolution of Biospheres*, 2005, 35, 29-60.

163. Tamura, M.; Kandori, R.; Kusakabe, N.; Nakajima, Y.; Hashimoto, J.; Nagashima, C.; Nagata, T.; Nagayama, T.; Kimura, H.; Yamamoto, T.; Hough, J. H.; Lucas, P.; Chrysostomou, A.; Bailey, J. *Astrophys J*, 2006, 649, 29-32.

164. Buschermöhle, M.; Whittet, D. C. B.; Chrysostomou, A.; Hough, J. H.; Lucas, P. W.; Adamson, A. J.; Whitney, B. A.; Wolff, M. J. *Astrophys J*, 2005, 624, 821-826.

165. Fukue, T.; Tamura, M.; Kandori, R.; Kusakabe, N.; Hough, J. H.; Lucas, P. W.; Bailey, J.; Whittet, D. C. B.; Nakajima, Y.; Hashimoto, J.; Nagata, T. *Astrophys J L*, 2009, 692, 88-91.